\newcommand{\pt}{p_{\rm T}}
\newcommand{\ptcut}{p_{\rm T, cut}}
\newcommand{\GeV}{\mathrm{GeV}}
\newcommand{\dndeta}{\mathrm{d}N_{\mathrm{ch}}/\mathrm{d}\eta}
\newcommand{\JPGNPP}{J.~Phys.~G:~Nucl.~Part.~Phys.~}
\newcommand{\INEL}{\rm{INEL}>0_{|\eta|<0.8, \pt>\ptcut}}
\documentclass{PoS}
\usepackage{lineno}
\usepackage{graphicx}
\usepackage{enumitem}
\title{Charged Particle Multiplicity and Pseudorapidity Density Measurements in pp collisions with ALICE at the LHC}

\ShortTitle{Charged Particle Multiplicity and Pseudorapidity Density Measurements in pp collisions with ALICE at the LHC}

\author{\speaker{Chiara Zampolli for the ALICE Collaboration}%
       INFN Bologna \& CERN\\
       E-mail: \email{chiara.zampolli@cern.ch}}

\abstract{These proceedings describe the charged-particle pseudorapidity densities and multiplicity distributions measured by the ALICE detector in 
pp collisions at $\sqrt{s} = 0.9$ and 7 TeV in specific phase space regions. The pseudorapidity range $|\eta| < 0.8$, together with $\pt$ cuts at 0.15, 0.5 and 1 $\GeV/c$ is considered. 
The classes of events considered are those having at least one charged particle in the kinematical ranges just described.
The results obtained by ALICE are compared to Monte Carlo predictions.
}

\FullConference{The European Physical Society Conference on High Energy Physics \\
		 18-24 July, 2013\\
		 Stockholm, Sweden}

\begin{document}

\section{Introduction}
\label{sec:Introduction}
The ALICE results 
on charged-particle pseudorapidity density ($\dndeta$) and multiplicity distributions in pp collisions at $\sqrt{s} = 0.9$ and 7 TeV presented in this document derive from an analysis 
carried out with an event and track selection specially chosen to make comparison with Monte Carlo calculations to allow for a better Monte Carlo tuning. 
Tracks reconstructed as coming from primary 
particles\footnote{The ensemble of primary charged particles includes those produced in the collision and their decay products, excluding weak decays from strange particles.} in the Inner Tracking
System (ITS) and in the Time Projection Chamber (TPC) of ALICE have been used, and kinematical phase space regions defined in $\eta$ ($|\eta| < 0.8$) and $\pt$ ($\pt$ $>$ $\ptcut$, with 
$\ptcut$ = 0.15, 0.5 and 1.0 $\GeV/c$) have been considered. The first $\ptcut$ corresponds to the $\pt$$_{\rm{cutoff}}$ at which the ALICE global
tracking efficiency (i.e. including both ITS and TPC) reaches $\sim 50\%$ and stays approximately constant ($\sim 70\-- 75\%$) for higher $\pt$~\cite{Aamodt:2010my}, 
while the higher $\ptcut$ values
allow a comparison with measurements performed by ATLAS and CMS. The pseudorapidity density and multiplicity distributions are measured for
all charged particles in the given $\pt-\eta$ region for those events which have 
at least one charged particle.
This introduces a so-called ``hadron level definition'' of the event class considered, named 
INEL$>0_{|\eta|<0.8, \pt>\ptcut}$ hereafter.

\section{The ALICE experiment and the data samples}
\label{sec:ALICEandData}
The ALICE experiment consists of a set of different detectors placed in a solenoidal magnetic field of 0.5 T (the central barrel) plus other detectors outside. 
Details about the various subsystems can be found in~\cite{ALICEdet}. For the 
analysis presented herein, tracks reconstructed in the ALICE central barrel by the ITS and TPC detectors were used, while the triggering and event selection relied on both the ITS and VZERO 
detectors.

The data samples used consisted of Minimum Bias pp events collected in 2009 and 2010. The Minimum Bias trigger was defined as a signal in either one of the two ALICE VZERO hodoscopes, 
or in the ITS pixel detector (one out of three). A coincidence with the signals from the two beam pick-up counters (BPTX) was also required to select the events and remove the background. 
In such conditions, about $110 000$ (collected in 2009)
and $2.2 \times 10^6$
events  (collected in 2010) were used for the charged-particle pseudorapidity density analysis at $\sqrt{s} = 0.9$ and 7 TeV respectively. The multiplicity distributions were obtained from approximately 
$2.9  \times 10^6$ and $2.7 \times 10^6$ events (all collected in 2010) at $\sqrt{s} = 0.9$ and 7 TeV respectively.

In addition to requiring the Minimum Bias trigger in the collision and the reconstruction of the primary vertex, a preselection of the events aimed
at reducing the beam background was applied based on the information from the VZERO detector and on the correlation between the number of hits and the so-called 
tracklets\footnote{A tracklet is built 
combining a pair of hits in the two innermost ITS layers.} found in the the two innermost ITS layers, corresponding to the Silicon Pixel Detector (SPD, 
see also~\cite{ALICEfirst}). A selection on the vertex was also applied,
requiring it to be obtained either from the tracks reconstructed from the TPC and the ITS detectors, or, in case this was not available, from the tracklets
found in the SPD detector (see also~\cite{ALICEfirst}), using the SPD information. 
Moreover, only events for which the vertex position along the $z$ coordinate ($vtx_{z}$) was such that $|vtx_{z}| < 10$ cm were accepted. Finally, only those events 
with at least one reconstructed track in the kinematical region defined by the pseudorapidity interval $|\eta| < 0.8$ and $\pt > \ptcut$ ($\ptcut = 0.15, 0.5, 1$ $\GeV/c$) were considered. 
For the 2010 data, an additional event selection criterion was used in order to reduce the contribution from pile-up events, removing those identified as coming from pile-up based on the SPD information. 
This sample was then corrected back to the $\INEL$ ``hadron level definition'' described in Sec.~\ref{sec:Introduction}.

\section{Analysis strategy}
\label{sec:AnalysisStrategy}
The tracks used in the analysis are those reconstructed by the ALICE central global tracking~\cite{PPR2}, which is based on the Kalman filter technique~\cite{PPR2,Kalman}. 
Track selection criteria (cuts) have been applied in order to maximize the tracking efficiency and minimize the contamination from secondaries and fake tracks, as described in~\cite{ALICE-PUB}.
The raw $\dndeta$ and multiplicity distributions obtained from data were corrected using PYTHIA Monte Carlo simulations as described in Sec.~\ref{sec:dndeta} and Sec.~\ref{sec:multiplicity}. 
The GEANT3 particle transport package was used together with a detailed 
description of the geometry of the experiment, and of the detector and electronics response. Moreover, the simulation was set to reproduce the conditions of the LHC beam and of the detectors
(in terms of vertex position, calibration, alignment and response) at the time the data under study were collected. 
The underestimate in the Monte Carlo simulations of the event strangeness content with respect to that found in the
data was also taken into account during the correction phase. Further 
information about the analysis strategy and the corrections applied can be found in~\cite{ALICE-PUB}.

\subsection{Corrections for the charged-particle pseudorapidity density analysis}
\label{sec:dndeta}
The charged-particle pseudorapidity distribution is given by the expression
\begin{equation}
\frac{1}{N_{\rm{ev}}} \frac{{\rm{d}}N_{\rm{ch}}}{\rm{d}\eta}.
\label{eq:dndeta}
\end{equation}
where $N_{\rm{ev}}$ corresponds to the total number of events that belong to the INEL$>0_{|\eta|<0.8, \pt>\ptcut}$ class.
The corrections applied to the raw $\dndeta$ distribution are of three types: a track-to-particle correction is needed, in order to take into account the difference between the measured 
tracks and the true charged primary particles coming from acceptance effects, detector and reconstruction efficiency; a second correction is applied,
to account for the fact that events without a 
reconstructed vertex 
are not considered; finally, the bias due to the INEL$>0_{|\eta|<0.8, \pt>\ptcut}$ event selection used is considered. 

\subsection{Corrections for multiplicity distribution analysis}
\label{sec:multiplicity} 
For the multiplicity analysis, the correction procedure is twofold. First, an unfolding technique is applied in order to account  for the fact that due to the efficiency, acceptance and 
detector effects, the measured multiplicity spectrum is distorted from the true one. In addition, vertex reconstruction and event selection efficiency need to be taken into consideration, 
in a similar way to the $\dndeta$ analysis.

The unfolding procedure used for the analysis presented here is described in~\cite{ALICE-PUB}. It is based on a $\chi^2$-minimization 
approach, where the $\chi^2$ function used to evaluate the ``guessed'' 
unfolded spectrum $U$, can be written as:
\begin{equation}
\hat{\chi}^2(U)=\sum_m \left ( \frac{M_m - \sum_t R_{mt}U_t}{e_m} \right )^2.
\end{equation}  
Here, $M_m$ is the measured distribution at true multiplicity $t$ with error $e_m$, and $R_{mt}$ is the response matrix element for measured multiplicity $m$ and true multiplicity $t$
which encodes the probability that an event with true multiplicity $t$ is measured with multiplicity $m$.
Because this minimization suffers from oscillations in the unfolded spectrum, a constraint $P(U)$ was added to the $\chi^2$ function, favouring a certain shape in the unfolded 
distribution~\cite{Blobel}, following the same approach as described and discussed in~\cite{ALICEsecond}.
The constraint $P(U)$ is called a $regularization$ $term$, and
the new function to be minimized becomes:
$$
\chi^2(U) = \hat{\chi}^2(U) + \beta P(U).
$$ 
As written in the formula, $P(U)$ depends only on the unfolded spectrum $U$. $\beta$ is the weight of the regularization term. 

For the unfolding procedure, a parameterization of the response matrix was used, in order to avoid statistics issues at high multiplicities and 
in the tails of the distributions for a given fixed true multiplicity. 
More details on the choice of the regularization function and on the parameterization
can be found in~\cite{ALICE-PUB}.

\subsection{Systematic Uncertainties}
\label{sec:SystematicUncertainties}
Various sources of systematic uncertainties have been taken into account, most of which are common to the two analyses. They include:
\begin{itemize}[noitemsep]
	\item track quality cuts variation;
	\item tracking efficiency;
	\item material budget;
	\item particle species relative fraction;
	\item event type (Single, Double, Non-Single Diffractive) relative composition;
	\item pile-up;
	\item Monte Carlo generator dependence of the corrections;
\end{itemize}	
Moreover, for the multiplicity distribution analysis only, the following sources of systematic uncertainties have been studied:
\begin{itemize}[noitemsep]
	\item choice of the regularization function and weight;
	\item bias introduced by the regularization~\cite{Cowan:2002in};
		\item unfolding dependence on the $\langle \pt \rangle$ as a function of multiplicity. 
\end{itemize}

The complete description of the 
evaluation of the systematic uncertainties is discussed in~\cite{ALICE-PUB}. 

%\vspace{1cm}
\section{Results}
\label{sec:Results}
\begin{figure}[t!]
\begin{center}
\begin{tabular}{cc}
\hspace{-0.7cm}
\includegraphics[height=7.5cm]{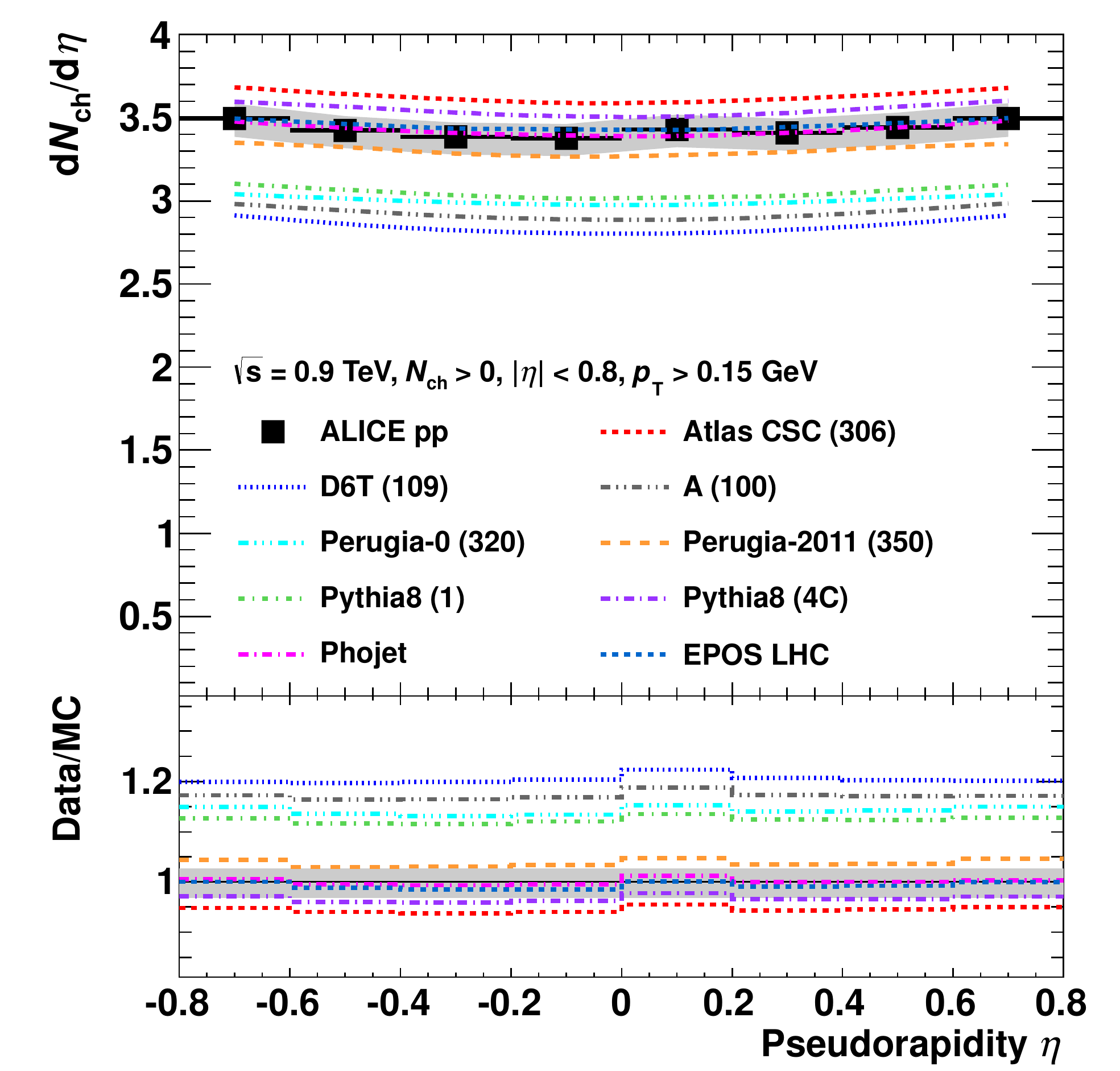} &
\hspace{-0.8cm}
\includegraphics[height=7.5cm]{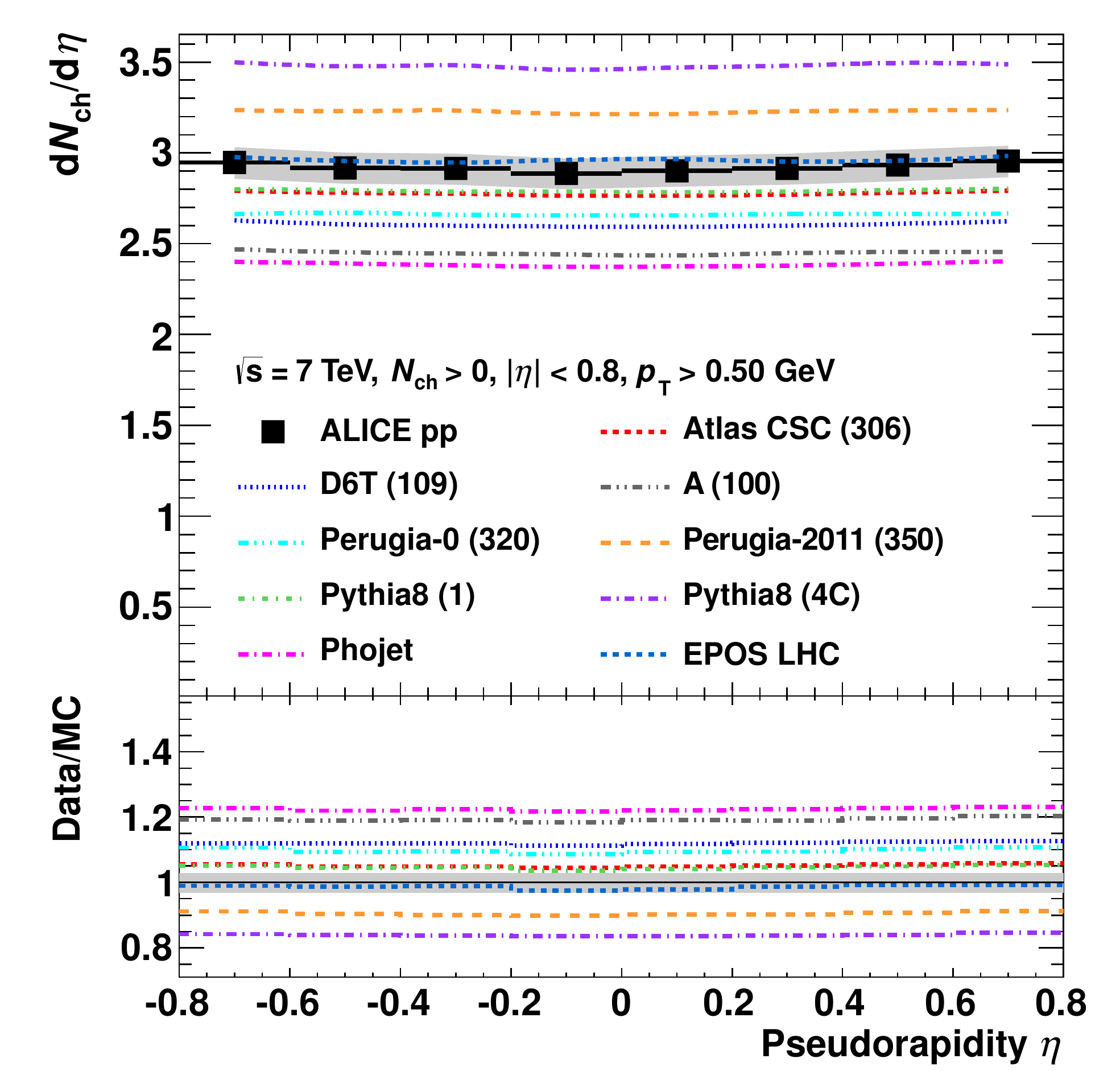} 
\end{tabular}
\end{center}
\vspace{-0.5cm}
\caption{$\dndeta$ versus $\eta$ obtained at $\sqrt{s} = 0.9$ (left) $\pt > 0.15$ $\GeV/c$ (left) and $\sqrt{s} = 7$ (left) $\pt > 0.5$ $\GeV/c$ for $|\eta| < 0.8$ normalized to the 
$\INEL$ event class. The predictions from different Monte Carlo generators are also shown.
The grey bands represent the systematic uncertainties on the data.
Bottom panels: data to Monte Carlo prediction ratios for the different generators considered. Here, the grey bands represent the total (statistical + systematic) uncertainty on the data.} 
\label{fig:dndeta}
\end{figure}

Figure~\ref{fig:dndeta} shows the final charged particle $\dndeta$ for the $\INEL$ classes of events. Here, for the sake of brevity, only the results for the $\ptcut = 0.15$ GeV/$c$ at $\sqrt{s} = 0.9$ (left panel) and $\ptcut = 0.5$ 
GeV/$c$ at 7 TeV (right) are shown. The complete set of results is presented in~\cite{ALICE-PUB}. 
Predictions from Monte Carlo generators are superimposed on the distributions. 
They are indicated as follows:
\begin{itemize}[noitemsep]
	\item PYTHIA-6
		\begin{itemize}[noitemsep]
			\item Atlas CSC (tune 306~\cite{CSC}); 
			\item D6T (tune 109~\cite{D6T}); 
			\item A (tune 100~\cite{A100});
			\item Perugia-0 (tune 320~\cite{Perugia0}); 
			\item Perugia-2011 (tune 350~\cite{Skands:2010ak}). 
		\end{itemize}
	\item PYTHIA-8
		\begin{itemize}[noitemsep]
			\item Pythia8 (tune 1~\cite{Pythia8});
			\item Pythia8 (tune 4C)~\cite{Corke:2010yf});
		\end{itemize}
	\item PHOJET (\cite{Phojet});
	\item EPOS LHC (\cite{EPOS}).
\end{itemize}	 
The bottom panels of the figures show the ratio between the 
data and the Monte Carlo predictions. 

\begin{figure}[t!]
\begin{center}
\begin{tabular}{cc}
\hspace{-0.5cm} \includegraphics[height=6.3cm]{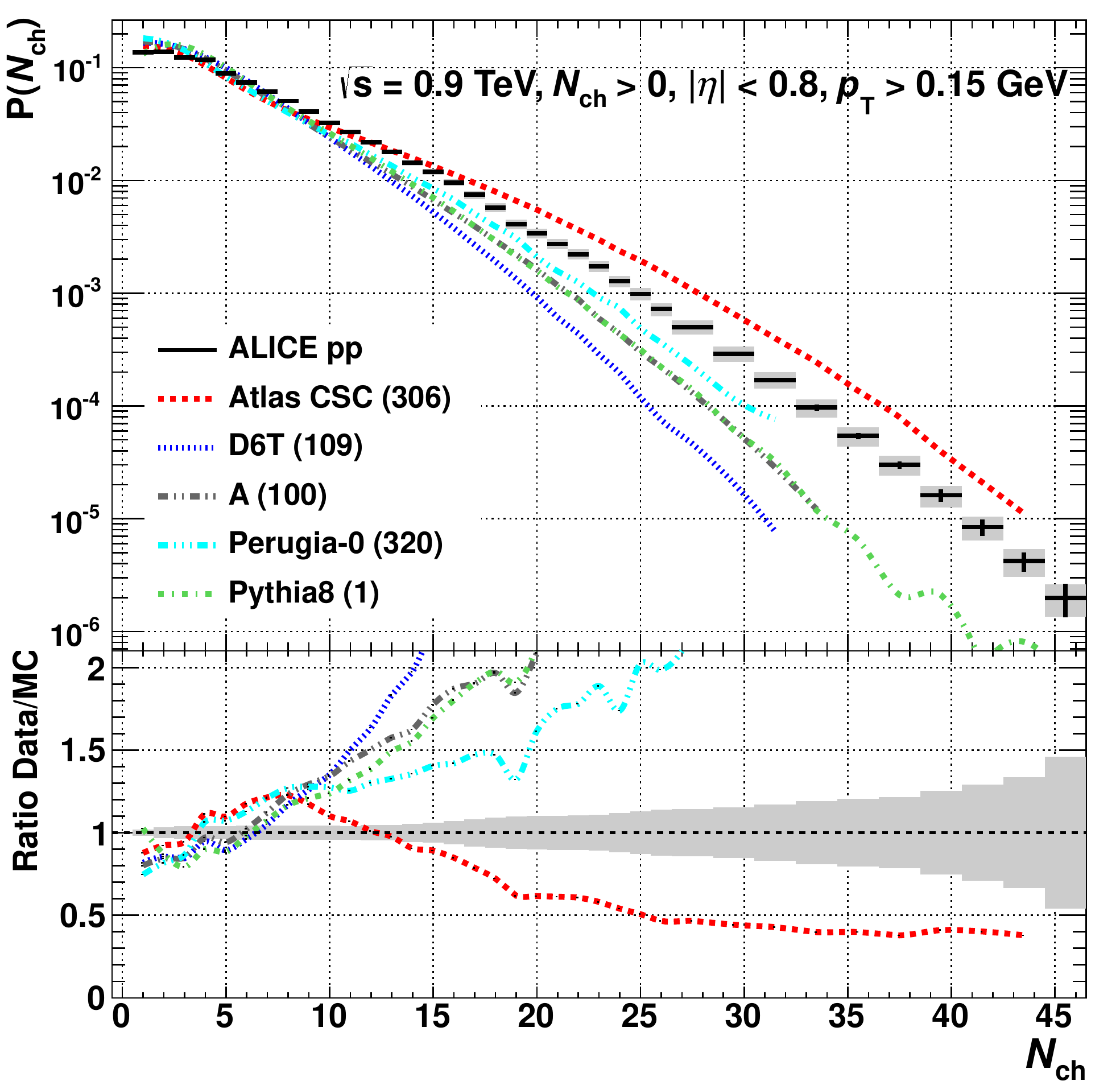} &
\hspace{-0.5cm} \includegraphics[height=6.3cm]{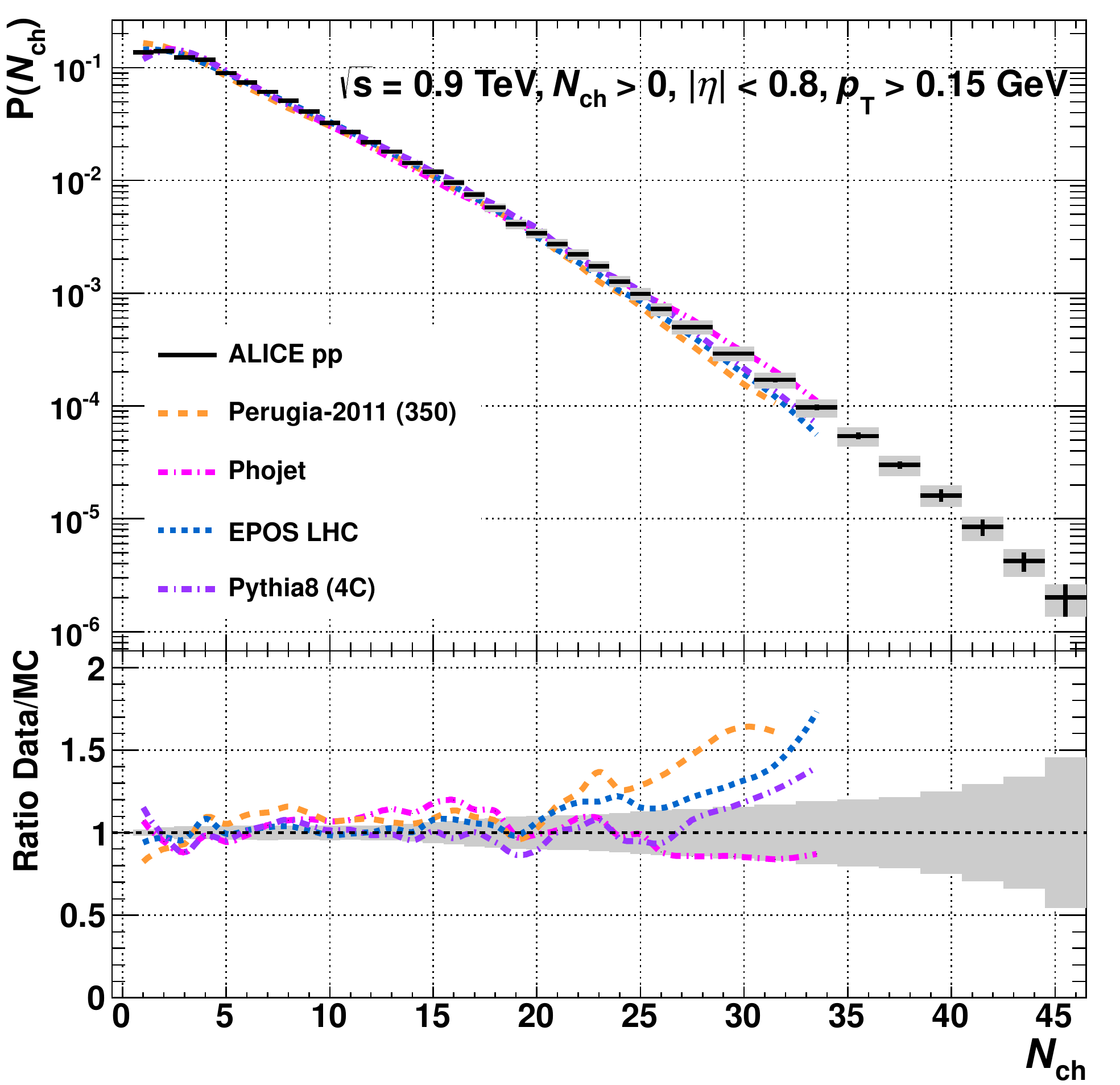} \\
\hspace{-0.5cm} \includegraphics[height=6.3cm]{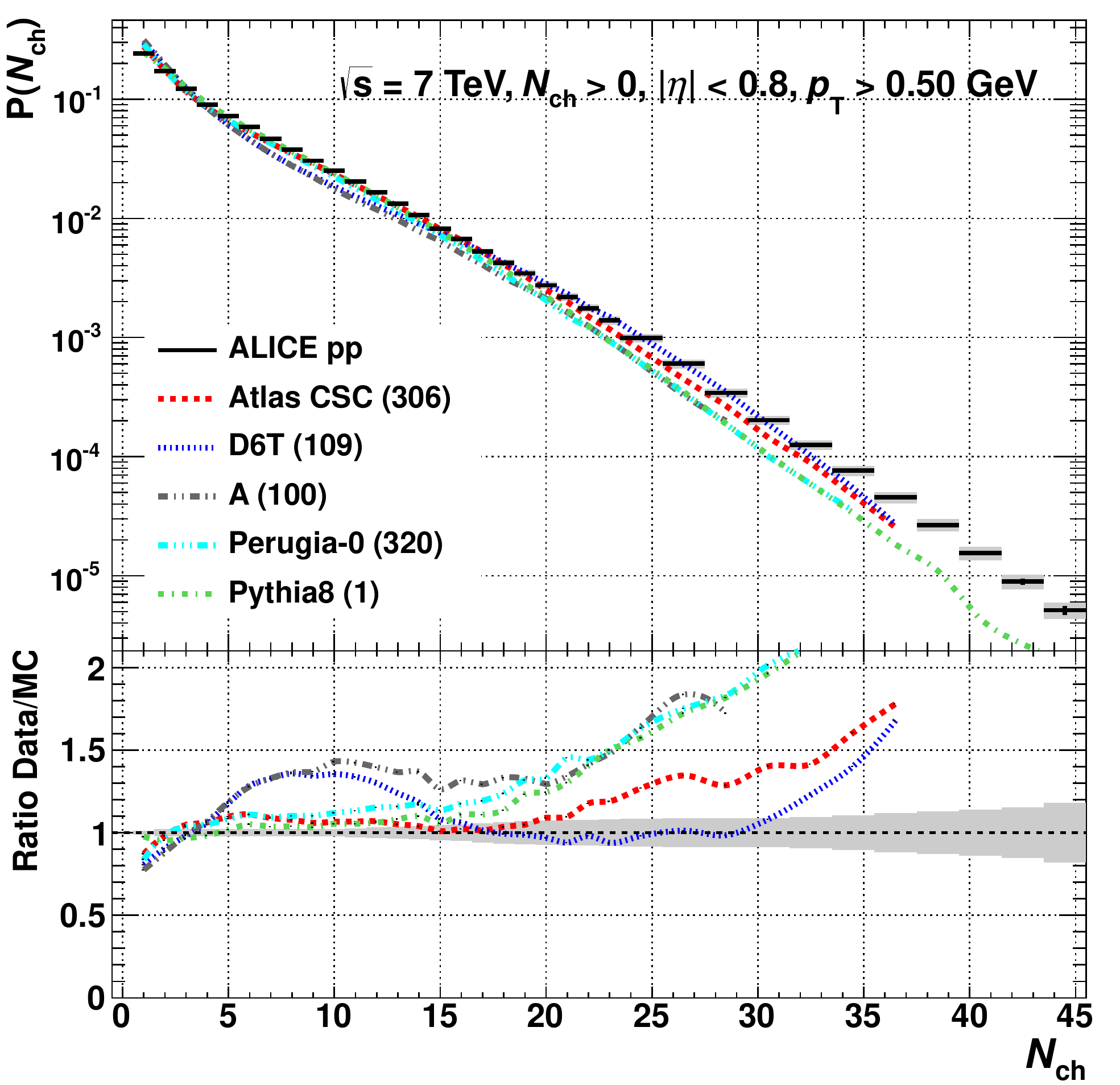} & 
\hspace{-0.5cm} \includegraphics[height=6.3cm]{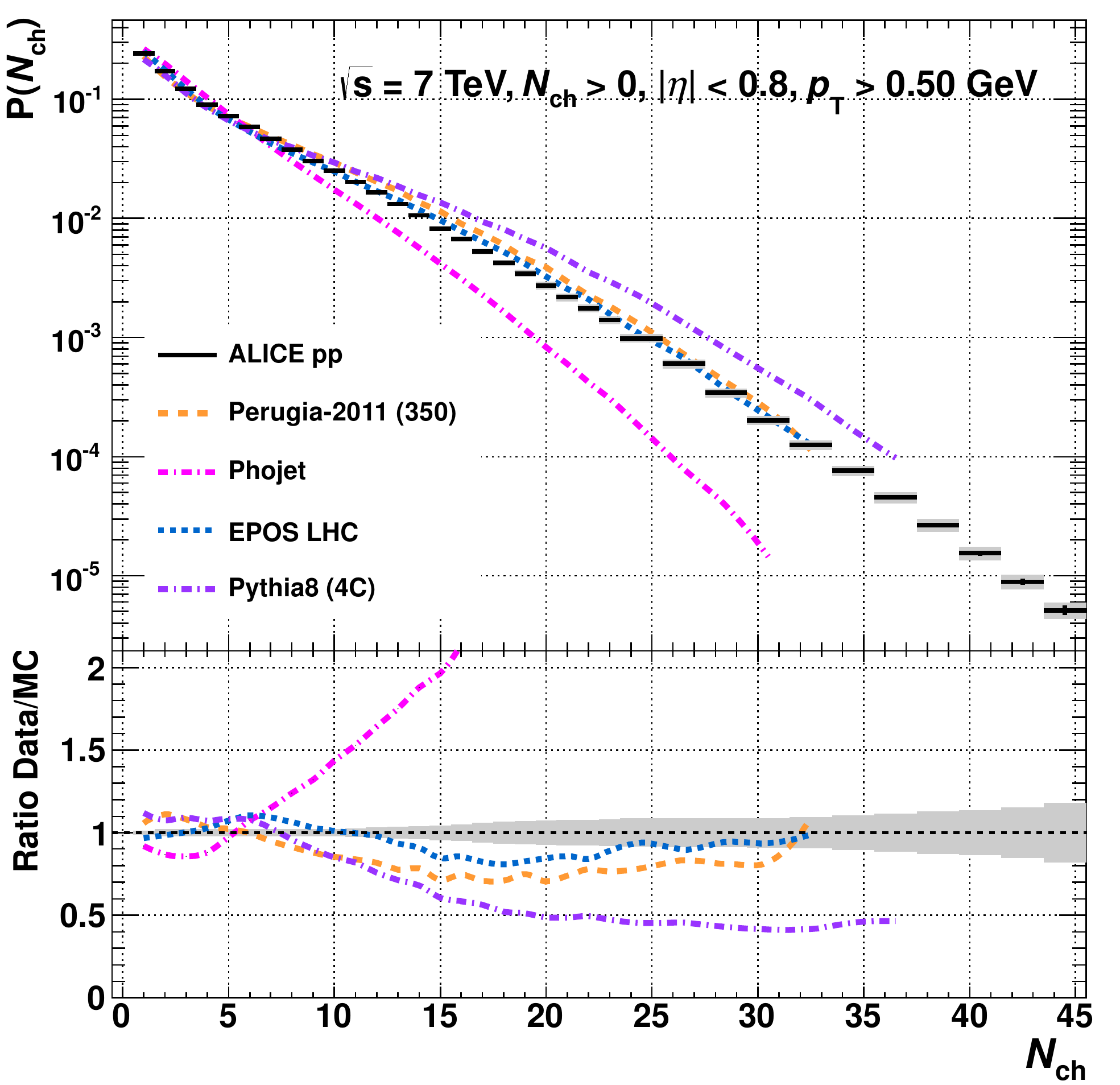} 
\end{tabular}
\end{center}
\vspace{-0.5cm}
\caption{Multiplicity distributions for the analysis at $\sqrt{s} = 0.9$ TeV, $\ptcut = 0.15$ GeV/$c$ (top) and at $\sqrt{s} = 7$ TeV, $\ptcut = 0.5$ GeV/$c$ (bottom). 
In the left and right panels, the data are compared with different Monte Carlo 
expectations, as indicated in the key. In the upper panels, 
data are shown with both statistical (black line) and systematic (grey band) uncertainties. The grey bands in the 
lower panels, where the data/Monte Carlo ratios are presented, correspond to
the total uncertainty on the final results.}
\label{fig:mult900}
\end{figure}

%\vspace{-1cm}
The charged particle multiplicity distribution results are shown in Fig.~\ref{fig:mult900} for the same cases as in Fig.~\ref{fig:dndeta}. For the 
sake of visibility, the Monte Carlo comparison (which was carried out with the same generators and tunes used for Fig.~\ref{fig:dndeta}) 
is shown for both figures in two different panels, as explained in the legends.

For both analyses, the results show that, in general, a universal trend in terms of comparison between the ALICE data and Monte Carlo calculations cannot be identified. 
At different centre-of-mass energies and with different values of $\ptcut$, the different models describe the data differently, and one tune that gives reasonable comparison to data in one case 
fails in the others. Moreover, for the 
multiplicity distributions, the level of agreement/disagreement varies significantly as a function of multiplicity.

\section{Conclusions}
\label{sec:Conclusions}
The charged particle pseudorapidity density and multiplicity distributions measured by ALICE at $\sqrt{s} = 0.9$ and 7 TeV
with charged tracks reconstructed in the ITS and TPC detectors have been presented. A $\ptcut$ (with $\ptcut$ = 0.15, 0.5, 1.0 GeV/$c$) was used in order to characterize the class of events to 
be considered for the analysis, namely the $\INEL$ class, defined 
requiring at least one charged particle with $p_t > \ptcut$ in $|\eta| < 0.8$. While the lowest $\ptcut$
allows the most inclusive measurement for ALICE  with global tracks, the 0.5 and 1.0 GeV$/c$ cuts were chosen together with the other LHC collaborations
(ATLAS, CMS) to allow for the comparison with their results (not shown here). The results were compared to different Monte Carlo models,
showing that the selected Monte Carlo generators do not reproduce the measurements at both centre-of-mass energies and for all choices of $\ptcut$.

\end{document}